# Electric Field Effect on Magnetization and Magnetocrystalline Anisotropy at the Fe/MgO(001) Interface

Manish K. Niranjan[1*], Chun-Gang Duan[2], Sitaram S. Jaswal[1] and Evgeny Y. Tsymbal[1†]

[1]*Department of Physics and Astronomy & Nebraska Center for Materials and Nanoscience, University of Nebraska, Lincoln, Nebraska 68588, USA*

[2]*Key Laboratory of Polarized Materials and Devices, Ministry of Education, East China Normal University, Shanghai 200062, China*

Density-functional calculations are performed to explore magnetoelectric effects originating from the influence of an external electric field on magnetic properties of the Fe/MgO(001) interface. It is shown that the effect on the interface magnetization and magnetocrystalline anisotropy can be substantially enhanced if the electric field is applied across a dielectric material with a large dielectric constant. In particular, we predict an enhancement of the interface magnetoelectric susceptibility by a factor of the dielectric constant of MgO over that of the free standing Fe (001) surface. We also predict a significant effect of electric field on the interface magnetocrystalline anisotropy due to the change in the relative occupancy of the 3d-orbitals of Fe atoms at the Fe/MgO interface. These results may be interesting for technological applications such as electrically controlled magnetic data storage.

Materials that have high magnetocrystalline anisotropy (MCA), especially nanometer-thick ferromagnetic films, are widely utilized in modern perpendicular magnetic recording technology (see, e.g., ref. 1). The large coercivity of such materials requires, however, high magnetic field to "write" bit information on them. In the past decade, various methods have been proposed to solve this problem, e.g., heat-assisted recording or the application of anisotropy-graded and exchange-coupled media. An alternative way to tailor the MCA (that has not yet been realized in practical devices) is to exploit the magnetoelectric (ME) effect.[2,3] The ME effect allows changing the bulk magnetization of a material via applying an electric field.[4] In a broader vision, ME effects also involve the electrically-controlled surface (interface) magnetization,[5-11] magnetic order,[12,13] exchange bias,[14-16] spin transport,[17-22] and magnetocrystalline anisotropy.[8,23-31] Since the MCA determines stable orientations of magnetization, the latter approach is especially promising – tailoring the magnetic anisotropy by electric fields may yield entirely new paradigms for magnetic data storage.

For metallic ferromagnets, electronically-driven ME effects are confined to the interface and originate from spin-dependent screening.[32] Consequently the electric field affects only the surface (interface) MCA. Experimentally, a change in the surface MCA of a few percent was observed in FePt(Pd)/electrolyte films.[23] Magnetic easy axis manipulation by electric field was also demonstrated in the dilute magnetic semiconductor (Ga,Mn)As.[25] Recently, a strong effect of electric field on the interface MCA was demonstrated for the Fe/MgO (001) [26,27] and FeCo/MgO (001) [31] interfaces. It was found that the application of a relatively small field 0.1 V/nm leads to a 40% change of MCA for Fe films.[26,27] First-principles calculations have been performed and shown that rather large fields are required to observe a sizable change in the surface magnetization and MCA.[8,28-30]

In this article we demonstrate that the electric field effect on the surface magnetization and MCA can be substantially enhanced if the electric field is applied across a dielectric material with a sufficiently large dielectric constant $\varepsilon$. Since the induced screening charge scales with the dielectric constant $\sigma = \varepsilon_0 \varepsilon E$, for high permittivity dielectrics the ME effect may be enhanced by orders of magnitude. To illustrate the significance of this prediction we explore the ME effect at the Fe/MgO (001) interface using density-functional calculations. We demonstrate the enhancement of the surface ME susceptibility by a factor of the dielectric constant of MgO over that of a free standing Fe (001) surface. We also find a significant increase in the electric field effect on the surface MCA due to the change in the relative occupancy of the 3d-orbitals of Fe atoms at the Fe/MgO interface.

To study the effect of electric field on the interface magnetization and MCA energy of the Fe/MgO (001) interface we perform density-functional calculations using a MgO/Fe/Cu(001)/Vacuum supercell. We employ the projected augmented wave (PAW) method [33] and the generalized gradient approximation (GGA) for exchange and correlation, as implemented within Vienna *Ab-Initio* Simulation Package (VASP).[34] We use standard plane wave basis set with a kinetic energy cutoff of 500 eV and the k-mesh sampled using 10×10×1 k-points in the full Brillouin zone. Along the [001] direction the supercell consists of nine monolayers of bcc Fe on top of nine monolayers of MgO followed by four monolayers of bcc copper and a vacuum layer. At the Fe/MgO interface, the O atoms are placed atop Fe atoms, consistent with the experimental data.[35] The electric field is introduced by the dipole layer method [36] with the dipole placed in the vacuum region of the supercell. The electric field points *away* from the Fe layer at Fe/MgO interface. The Cu layer is used to eliminate the screening charge at the otherwise Fe/Vacuum interface. The in-plane lattice constant of the supercell is kept fixed at the experimental lattice constant of Fe ($a$ = 2.87 Å). The structures are relaxed in the absence of the electric field until the largest force becomes less than 5.0 meV/Å. The MCA energy is obtained by taking the difference between the total energy calculated within the force theorem corresponding to magnetization pointing along the [100] and [001] directions in the presence of spin-orbit interaction.

First, we consider the effect of electric field on the magnetization at the Fe/MgO interface. Here and below the ME effects are discussed as a function of the electric field *within* the MgO layer relevant to experiments where the field is applied between two metal electrodes across the dielectric. Our calculations find that the electric field in MgO is reduced by a factor of 3.1 as compared to that in vacuum due to dielectric screening. In the absence of ionic relaxations this reduction is entirely caused by the electronic contribution to dielectric susceptibility of MgO which is associated with the high frequency dielectric constant $\varepsilon_\infty$. The calculated value of $\varepsilon_\infty \approx 3.1$ is in agreement with the experimental value $\varepsilon_\infty \approx 3.0$.[37]

The electric field is screened by free charges in Fe at the Fe/MgO interface. Due to ferromagnetism of Fe this screening is spin-dependent and leads to the induced interface magnetization, resulting in the surface (interface) ME effect.[32] The induced interface magnetization is evident from Fig. 1a which shows the change in the spin density $\Delta\sigma = \sigma(E) - \sigma(0)$ due to electric field $E = 1$ V/nm projected on the $x$-$z$ or (010) plane of the supercell. It is seen that only the spin density at the interfacial Fe atoms is changed significantly. This result is qualitatively similar to that found for the Fe(001) surface.[8]

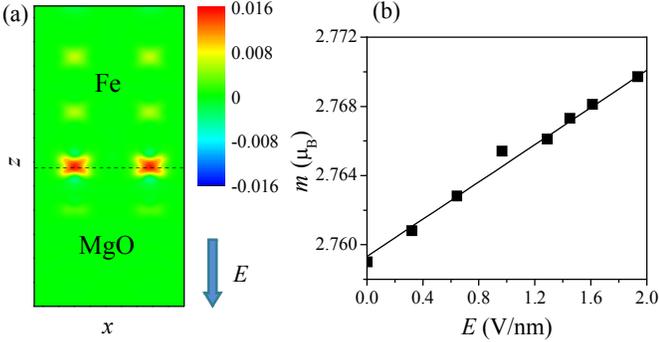

**Fig. 1**: (a) Induced spin density $\Delta\sigma = \sigma(E) - \sigma(0)$, in units of $e/\text{Å}^3$, projected to the $x$-$z$ or (010) plane around the Fe/MgO interface under the influence of electric field $E = 1.0$ V/nm in MgO. The dashed line indicates the interfacial Fe monolayer at the Fe/MgO interface. (b) Magnetic moment (in units of $\mu_B$) of Fe at Fe/MgO interface as a function of the electric field in the MgO.

Fig. 1b shows the magnetic moment $m$ of the Fe atom at the Fe/MgO interface versus the electric field $E$ in MgO. It is seen that within the computational error $m$ changes linearly with $E$. The induced interface magnetization $\Delta M$ is given by $\mu_0 \Delta M = \alpha_S E$, where $\alpha_S$ is the surface (interface) ME coefficient.[8] We find that $\alpha_S \approx 1.1 \times 10^{-13}$ Gcm$^2$/V. This value is larger than that for the Fe(001) surface [8] by a factor of 3.8, which is approximately equal to the calculated $\varepsilon_\infty$. The larger $\alpha_S$ is due to the enhancement of the screening charge at the Fe/MgO interface. Within a rigid band model the surface ME coefficient is given by $\alpha_s = \frac{\varepsilon \mu_B}{ec^2} P$,[8] where $P$ is the spin polarization of the interface density of states at the Fermi energy. This result indicates that $\alpha_S$ is scaled linearly with $\varepsilon$ due to the enhanced surface charge density $\sigma = \varepsilon_0 \varepsilon E$ unequally distributed between the interface majority- and minority-spin states.

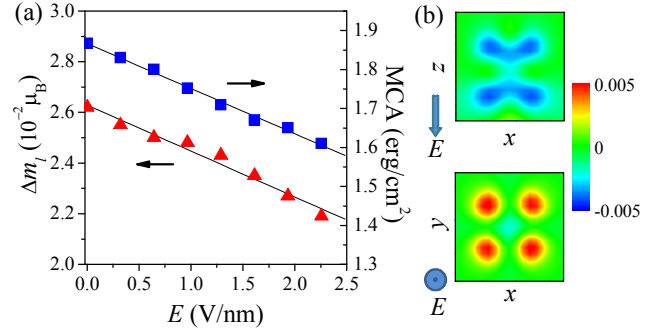

**Fig. 2**: (a) Magnetic anisotropy energy (MCA, squares) and orbital moment anisotropy ($\Delta m_l$, triangles) of the Fe/MgO(001) interface as a function of electric field in MgO. (b) Induced charge density $\Delta\rho = \rho(E) - \rho(0)$, in units of $e/\text{Å}^3$, at the interfacial Fe atom for $E = 1.0$ V/nm in the $x$-$z$ (010) plane (top panel) and the $x$-$y$ (001) plane (bottom panel).

Next we discuss the electrically induced MCA at the Fe/MgO (001) interface. For a given $E$ we calculate MCA of the MgO/Fe/Cu/Vacuum structure and subtract the MCA value of the Fe/Cu interface obtained in a separate calculation for a Fe/Cu(001) supercell. Since the contribution of the Fe/Cu interface to the total MCA energy does not depend on $E$, due to the screening charge being confined at the Cu surface, this allows us to obtain the Fe/MgO(001) interface MCA energy as a function of $E$. The results are shown in Fig. 2a. It is seen that within the accuracy of our calculation the MCA of the Fe/MgO interface changes linearly with electric field. We can define the surface (interface) MCA coefficient $\beta_S$ according to $\Delta K = \beta_S E$, where $\Delta K$ is the change in the MCA energy. From the slope in Fig. 2a, we find that the surface MCA is changed by 0.10 erg/cm$^2$ for $E = 1$ V/nm resulting in $\beta_S \approx 10^{-8}$ erg/Vcm. This value is larger by a factor of 5 than that obtained for the Fe(001) surface, i. e. $\beta_S \approx 2 \times 10^{-9}$ erg/Vcm.[8]

It is instructive to compare the variation in the MCA and the orbital moment anisotropy $\Delta m_l = m_l[001] - m_l[100]$, since they both arise from the spin-orbit interaction. Fig. 2a indicates that the two anisotropies are linearly related which is possible only if the majority $d$ band is fully occupied so that spin-flip matrix elements of the spin-orbit interaction can be neglected.[38] This fact allows us to discuss the behavior of MCA energy in terms of the orbital magnetic moment.

The decrease in MCA at the Fe/MgO (001) interface with electric field pointing away from the Fe layer originates from the redistribution of electron charge between different $d$-orbitals. Fig. 2b shows the change in the charge density, $\Delta\rho = \rho(E) - \rho(0)$, at the interface Fe atom induced by electric field $E = 1$ V/nm. From the $x$-$z$ (010) plane and the $x$-$y$ (001) plane projections we see a reduced occupation of the $d_{xz}$ (and by symmetry $d_{yz}$) orbitals (top panel in Fig. 2b) and the enhanced occupation of the $d_{xy}$ orbitals (bottom panel in Fig. 2b). The $L_z$ operator has positive non-zero matrix elements $\langle d_{xz} | L_z | d_{yz} \rangle$, while $\langle d_{xz} | L_z | d_{xy} \rangle = \langle d_{yz} | L_z | d_{xy} \rangle = 0$. Thus, the reduced occupation of the $d_{xz}$ and $d_{yz}$ orbitals leads to the reduction of $\langle L_z \rangle$, and hence a decrease in the associated orbital magnetic moment $m_l[001]$. The latter is evident from the calculated $m_l$[001] shown in Fig. 3a (squares), indicating a reduction of $m_l$[001] with electric field. On the other hand, as seen from Fig. 3a (triangles), $m_l$[100] is weakly dependent on $E$. Apparently the reduced occupation of the $d_{xz}$ ($d_{yz}$) orbitals and enhanced occupation of the $d_{xy}$ orbitals (Fig. 2b) are largely canceled out in the contribution to $\langle L_x \rangle$ (and hence to $m_l$[100]) through non-vanishing matrix elements $\langle d_{xz}|L_x|d_{xy}\rangle$ and $\langle d_{yz}|L_x|d_{xy}\rangle$. As pointed out earlier and shown in Fig. 2a, the difference of the two orbital moments and hence MCA decrease with increasing electric field. This behavior is opposite to that found previously for the Fe (001) surface.[8] For the latter, as seen from Fig. 3b, the two orbital magnetic moments diverge with increasing electric field (pointing away from Fe) resulting in the increase of the MCA energy.

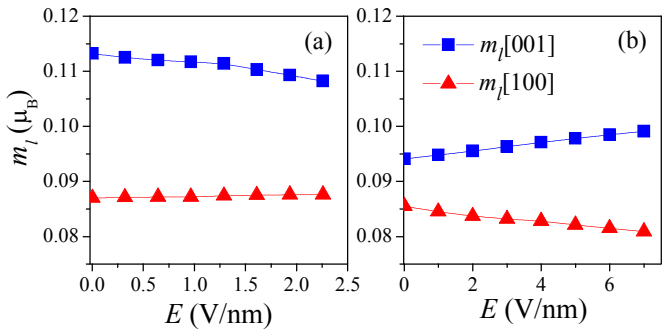

**Fig. 3**: Orbital magnetic moment of Fe atom at (a) the Fe/MgO (001) interface and (b) the Fe(001) surface, as a function of electric field, for the magnetic moment pointing along the [001] direction (squares) and along the [100] direction (triangles).

We have explored the effect of ionic relaxations driven by electric field. The calculation of the static dielectric constant of MgO in fields ranging from 0 to 3 V/nm predicts the average value of $\varepsilon \sim 8.5$ in agreement with the experimental value of $\varepsilon \approx 9.5$. For a given field in vacuum, changes in the interface magnetization and MCA are qualitatively similar to those for the unrelaxed structure. This is because the macroscopic screening charge primarily responsible for the ME effects depends only on the vacuum field due to the cancellation of the dielectric effects of the insulator. Since the electric field in MgO is scaled with its dielectric constant, this result implies that the ME effects for the ideal Fe/MgO(001) interface seen in static measurements should be proportionally enhanced. We note however that the interface MCA is very sensitive to the interface electronic and atomic structure. In particular, we find that MCA of the Fe/MgO(001) interface with additional O adsorbed in the Fe interfacial monolayer [35] becomes negative (about –0.7 erg/cm$^2$), hence suggesting an in-plane anisotropy. Thus the amount of oxygen at the Fe/MgO interface may significantly influence the interface MCA, which may explain much smaller values of the Fe/MgO interface anisotropy observed experimentally. [31] On the other hand, there are indications that the interface MCA of Fe/MgO (001) interfaces with controlled oxygen stoichiometry may exceed 1 erg/cm$^2$ consistent with our calculations. [39] These experimental results in conjunction with our predictions are very exciting in view of using Fe/MgO interfaces to control the interface magnetic anisotropy by electric fields. We therefore hope that our results will stimulate further experiments in this field.

This work was supported by the Nebraska MRSEC (NSF Grant No. 0820521), the Nanoelectronics Research Initiative, NSFC (Grant No. 50771072 and 50832003), and the Nebraska Research Initiative. Computations were performed utilizing the Research Computing Facility of University of Nebraska.

* E-mail: mniranjan2@unl.edu
† E-mail: tsymbal@unl.edu